\documentclass[prc,aps,showpacs,twocolumn,nofootinbib,superscriptaddress,floatfix]{revtex4-1}

\usepackage{graphicx}
\usepackage{epsfig}
\usepackage{amsfonts}
\usepackage{amsmath,amssymb}
\usepackage{bm}
\begin{document}

\title{Fission modes of mercury isotopes}%

\author{M. Warda}

\affiliation{Institute of Physics, Maria Curie-Sk{\l}odowska University, pl. M. Curie-Sk{\l}odowskiej 1, 20-031 Lublin, Poland}

\author{A. Staszczak}

\affiliation{Institute of Physics, Maria Curie-Sk{\l}odowska University, pl. M. Curie-Sk{\l}odowskiej 1, 20-031 Lublin, Poland}
\affiliation{
Department of Physics and Astronomy, University of Tennessee, Knoxville, Tennessee 37996, USA
}%
\affiliation{
Physics Division, Oak Ridge National Laboratory, Oak Ridge, Tennessee 37831, USA
}%

\author{W. Nazarewicz}
\affiliation{
Department of Physics and Astronomy, University of Tennessee, Knoxville, Tennessee 37996, USA
}%
\affiliation{
Physics Division, Oak Ridge National Laboratory, Oak Ridge, Tennessee 37831, USA
}%
\affiliation{
Institute of Theoretical Physics, University of Warsaw, ul. Ho\.za 69,
PL-00-681 Warsaw, Poland }%

%Background, Purpose, Methods, Results, and Conclusions

%date{\today}

\begin{abstract}
\begin{description}
\item[Background]
Recent experiments on beta-delayed fission in the mercury-lead region and the discovery of asymmetric fission in $^{180}$Hg \cite{And10}  have stimulated  theoretical interest in the mechanism of fission in heavy nuclei.
\item[Purpose]
We study fission modes and fusion valleys in   $^{180}$Hg and  $^{198}$Hg to 
reveal the role of shell effects in pre-scission region and
explain the experimentally observed fragment mass asymmetry and its variation with $A$.
\item[Methods]
We   use the self-consistent nuclear density functional theory employing Skyrme and Gogny energy density functionals.
\item[Results] 
The potential energy surfaces in multi-dimensional space of collective coordinates, including elongation, triaxiality, reflection-asymmetry, and necking,  are calculated for $^{180}$Hg and  $^{198}$Hg. 
The asymmetric fission valleys --  well separated from  fusion valleys associated with nearly spherical fragments -- are found in in both cases.
The density distributions at scission configurations are studied and related to  the experimentally observed  mass splits.
\item[Conclusions]
The energy density functionals SkM$^*$ and D1S give a very consistent description of the fission process in $^{180}$Hg and $^{198}$Hg.
We predict a transition from asymmetric fission in $^{180}$Hg towards more symmetric distribution of fission fragments in $^{198}$Hg.
For $^{180}$Hg, both models yield  $^{100}$Ru/$^{80}$Kr as the most probable split. For  $^{198}$Hg, the most likely split  is  $^{108}$Ru/$^{90}$Kr in HFB-D1S and $^{110}$Ru/$^{88}$Kr in HFB-SkM$^*$.

\end{description}
\end{abstract} 
%Recent experiments on beta-delayed fission in the mercury-lead region and the discovery of asymmetric fission in $^{180}$Hg \cite{And10} have stimulated renewed interest in the mechanism of fission in heavy nuclei. Here we study fission modes and fusion valleys in   $^{180}$Hg and  $^{198}$Hg  using the self-consistent nuclear density functional theory employing Skyrme and Gogny energy density functionals.
%We  show that the observed transition from asymmetric fission in $^{180}$Hg towards more symmetric distribution of fission fragments in $^{198}$Hg  can be explained in terms of competing fission modes of different geometries that are governed by shell effects in pre-scission configurations. The density distributions at scission configurations are studied and related to  the experimentally observed  mass splits.


\maketitle

%%%%%%%%%%%%%%%%%%%%%%%%%%%%%%%%%%%%%%%%%%%%%%%%%%%%%%%%
\section{\label{sec:intro}Introduction}

%\textit{Introduction}---
The fission phenomenon is a magnificent example of a quantal large-amplitude collective motion during which the nucleus evolves in a multidimensional space representing shapes with different geometries, often tunneling through a classically-forbidden region \cite{Kra12}.
Understanding the fission process is crucial for many areas of science and technology. For instance, fission governs the existence of many transuranium elements, including the predicted long-lived super-heavy species. In nuclear astrophysics, fission influences the formation of heavy elements in a very high neutron density environment. Fission applications are numerous. For instance, improved understanding of the fission process will enable scientists to enhance the safety and reliability of  nuclear reactors. While in the past the design, construction, and operation of reactors were supported through empirical trials, the new phase in nuclear energy production is expected to rely heavily on advanced modeling and simulation capabilities utilizing massively parallel leadership-class computers.

A comprehensive  explanation of nuclear fission rooted in interactions between nucleons still eludes us, although self-consistent approaches based on the nuclear density functional theory (DFT)
have recently demonstrated that a microscopic description has a potential for both qualitative and  quantitative description of fission data \cite{War02,Gou05,Sta09,war05,You09,Kor12}.
A starting point in the adiabatic approach to fission is the capability to compute accurate
multidimensional potential energy surfaces (PES), and use them to predict
observables such as fission half-lives and fragment mass distributions.

This work has been stimulated by recent experiments on beta-delayed fission in the mercury-lead region \cite{And10} and the discovery of asymmetric fission of the nucleus $^{180}$Hg. Such an outcome has not been  initially anticipated, as the symmetric fission channel involving two semi-magic $^{90}$Zr fragments was believed to dominate the  process. It has been generally expected that the asymmetric fission is not important  below $^{227}$Th \cite{Schmidt2000}, in particular in  pre-actinide nuclei with high-lying  saddle-point configurations that depend weakly on shell effects \cite{Myers1991}. Moreover, the data on  mass distributions of fragments in the low-energy fission of nuclei with $187\le A \le 213$  have demonstrated the strong presence of the symmetric fission mode \cite{Mul98}.
In particular, the nucleus  $^{198}$Hg has been observed to exhibit a fairly broad mass distribution \cite{Mul98,Itk90,*Itk91}.

The explanation of the asymmetric fission around $^{180}$Hg has been offered by the macroscopic-microscopic model \cite{And10,ich12} and its
extension \cite{Mol12} in terms of
an asymmetric fission pathway
that is separated by a potential-energy  ridge from the symmetric $^{90}$Zr+$^{90}$Zr fusion valley. These results have emphasized  the importance of shell effects between fission saddle and scission  in pre-actinide nuclei (see also Ref.~\cite{Kar08}).
\begin{figure*}[htb]
  \includegraphics[width=0.8\textwidth]{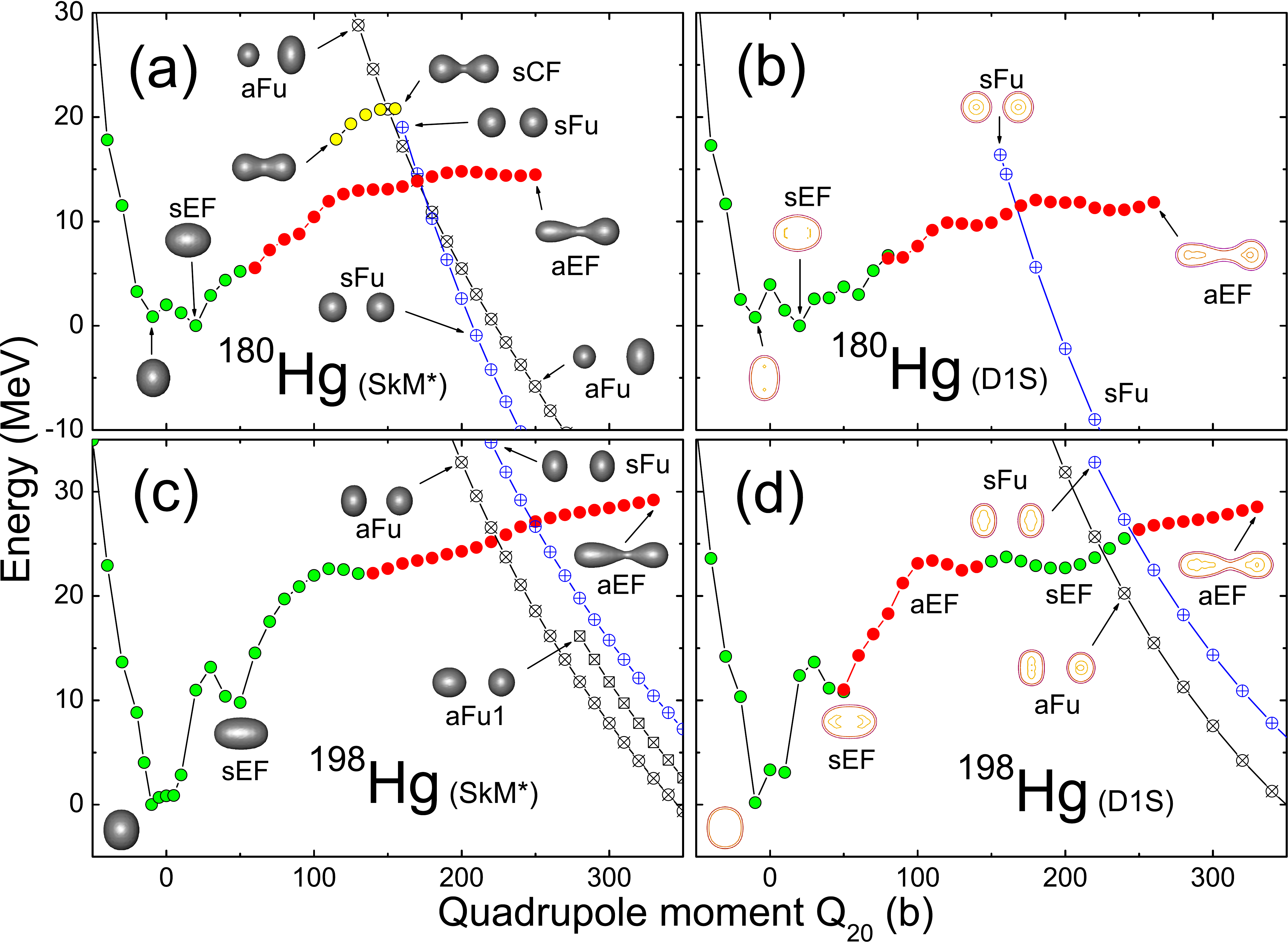}
  \caption[T]{\label{fig1}
(Color online)  Fission pathways for  $^{180}$Hg (top)  and $^{198}$Hg (bottom)  as functions of the driving quadrupole moment,
$Q_{20}$ calculated with  HFB-SkM$^*$  (left)  and HFB-D1S (right).
Competing fission
and fusion valleys are indicated, together with  the associate shapes. 
Since the sEF configuration sometimes corresponds to a ridge (rather than a valley), the corresponding curves do not continue  in such cases.
See text for details.
}
\end{figure*}
In this work, we extend the theoretical analysis of Ref.~\cite{And10} using  the  self-consistent nuclear DFT. We explain the transition from the asymmetric fission in
$^{180}$Hg \cite{And10} to a more symmetric situation in $^{198}$Hg \cite{Mul98} in terms of shell effects. We  compare the fission pathways  in both nuclei and discuss the interplay between  fission and fusion valleys. Finally, by studying density  distributions of fragments, we  demonstrate that scission configurations of $^{180}$Hg  and $^{198}$Hg   can be understood in terms of molecular structures.

%%%%%%%%%%%%%%%%%%%%%%%%%%%%%%%%%%%%%%%%%%%%%%%%%%%%%%%%
\section{\label{sec:model}Model}

%\textit{Model}---
The  Skyrme-HFB calculations were carried out using the
framework previously discussed in Refs.~\cite{Sta09,Sta10,Sta11}  based on the
symmetry unrestricted DFT solver HFODD \cite{Sch11} capable of breaking all self-consistent symmetries of nuclear mean fields on the way to fission.    To  solve a constrained
nonlinear HFB problem  precisely, we employed the Augmented Lagrangian method
\cite{Sta10}. The nuclear energy density functional was
approximated by the SkM$^*$ functional~\cite{Bar82} in the particle-hole channel and the
density-dependent mixed pairing interaction~\cite{Dob02} in the particle-particle channel. To truncate  the quasi-particle space of HFB,
we adopted the quasiparticle-cut-off value of 60\,MeV in the equivalent energy spectrum \cite{Dob84}. The pairing  strengths
were adjusted
to reproduce the neutron and proton pairing gaps in
$^{252}$Fm \cite{Sta09}; the resulting values are  $V_{n0}=-268.9$\,MeV\,fm$^3$  and $V_{p0}=-332.5$\,MeV\,fm$^3$.
The stretched  harmonic oscillator (HO) basis of HFODD was composed of states having not more than $N_{0}=26$ quanta in either of the
Cartesian directions, and not more than 1140 states in total.

Our Gogny calculations follow the framework described in Refs.~\cite{War02,War11} based on the
axial Gogny-HFB solver~\cite{axialcode,*Egi97,*rob11} and D1S parameter set \cite{ber84}. We used the  stretched HO
basis  with $N_z=22$ HO shells along the symmetry axis and $N_\perp=15$  shells in the perpendicular direction.
The oscillator length was adjusted at every calculation point.

To find the optimum  trajectories in a multidimensional
collective space, we  constrain the nuclear collective coordinates
associated with  the
 multipole moments $Q_{\lambda\mu}$, by which we explore the main
degrees of freedom related to elongation $(\lambda\mu=20)$ and
reflection-asymmetry $(\lambda\mu=30)$. In our symmetry unrestricted Skyrme-HFB calculations we also explore the effects of
triaxiality ($\lambda\mu=22$) and necking $(\lambda\mu=40)$. In our axial Gogny calculations, the scission configurations were studied by means of the
neck coordinate $Q_N$ defined through the gaussian-type operator
$\hat Q_N= \exp[-(z-z_0)^2/a^2]$ with $a=0.1$\,fm and $z_0$ chosen to describe the neck region (e.g., $z_0=0.5$\,fm in $^{180}$Hg).
$Q_N$  describes the number of nucleons in a thin layer of thickness $a$ perpendicular
to the symmetry axis placed at
$z = z_0$. Large values of $Q_N$  describe
shapes with a thick neck. By decreasing $Q_N$ one can approach  the scission line.
To obtain a PES, constrained HFB equations are solved to minimize  the total energy of the
system at each point in the collective space. As demonstrated earlier \cite{Sta09,war05}, exploring many collective coordinates makes it possible to identify saddle points~\cite{Mol01,*Dub11} as  the
competing  fission pathways are usually  well separated when studied in
more than one dimension.

%%%%%%%%%%%%%%%%%%%%%%%%%%%%%%%%%%%%%%%%%%%%%%%%%%%%%%%%
\section{\label{sec:fission}Competing fission modes}
\begin{figure}[htb]
  \includegraphics[width=\columnwidth]{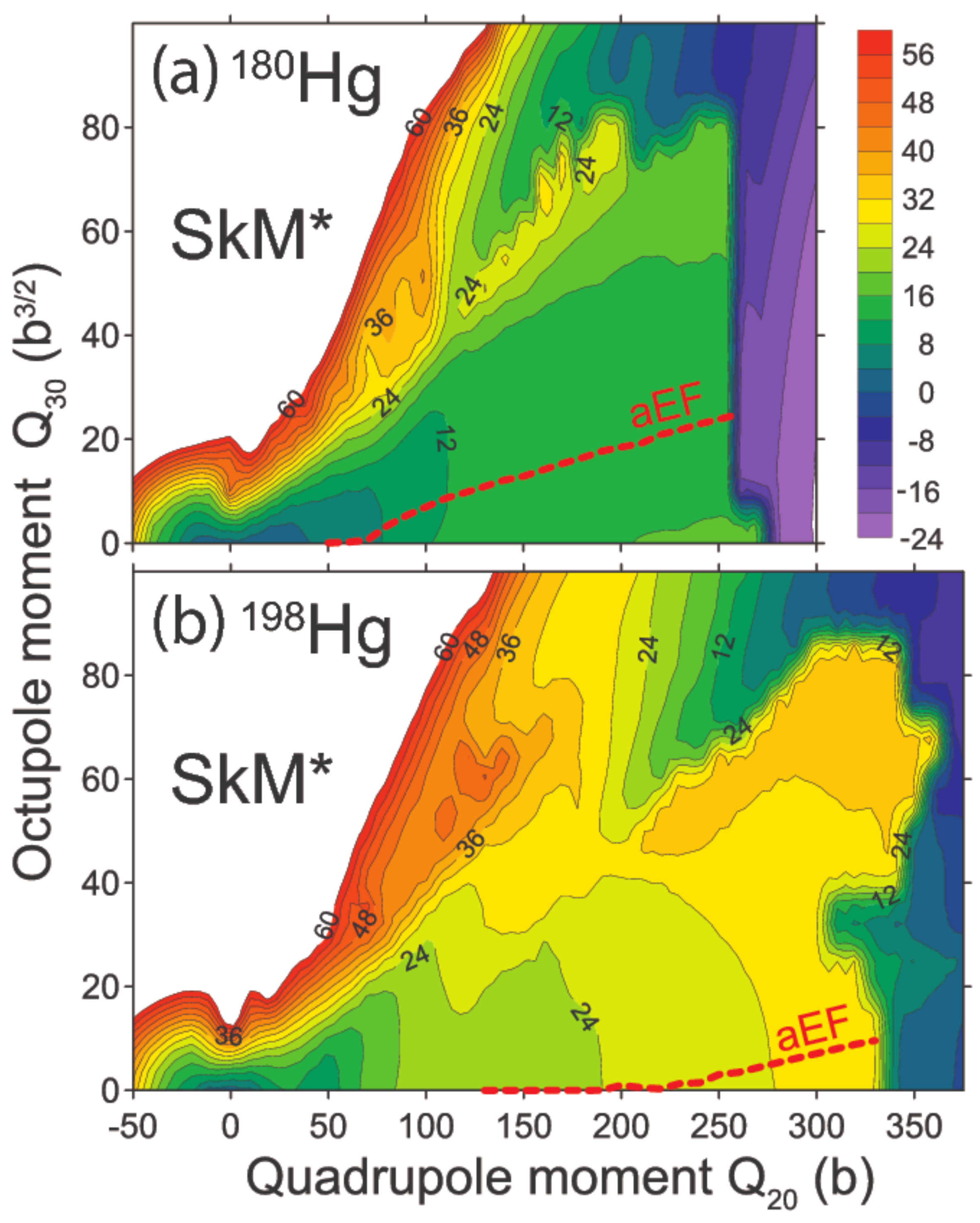}
  \caption[T]{\label{fig2}
  (Color online) PES
for $^{180}$Hg (top)  and $^{198}$Hg (bottom)   in the plane of collective coordinates $Q_{20}-Q_{30}$ in HFB-SkM$^*$.
The aEF fission pathway corresponding to  asymmetric elongated fragments  is marked.
The difference between contour lines is 4\,MeV. The effects due to triaxiality, known to impact inner fission barriers in the actinides, are negligible here.
 }
\end{figure}

%\textit{Competing fission modes}---
Figure~\ref{fig1} shows the calculated  fission pathways  in  $^{180}$Hg and $^{198}$Hg. Both models predict a fairly similar  pattern. 
The reflection symmetric fission path associated with associated with  elongated fragments (sEF) can be found for small deformations.  The reflection-asymmetric path, corresponding to elongated fission fragments (aEF) of different masses and shapes, is branching away from the symmetric valley below $Q_{20}=100$\,b,  and  it passes through the mass asymmetric scission point (see Fig.~\ref{fig2} for a better view of aEF in the $Q_{20}-Q_{30}$ plane). 

In $^{180}$Hg, at large elongations, the aEF path is strongly favored over reflection-symmetric configurations  ($Q_{30}=0)$ associated with  elongated fragments (sEF)
and the  symmetric compact fragment valley sCF -- marked in Fig.~\ref{fig1}(a).
The fusion valleys, both symmetric (sFu) and asymmetric (aFu) appear very low in energy above $Q_{20}\approx 200$\,b; they are associated with post-scission configurations in which the two fragments are well separated. Since the fission process is adiabatic, sFu and aFu are not expected to couple to
aEF, which has a very distinct compound configuration exhibiting a pronounced neck even at $Q_{20}\approx 250$\,b.

The situation in $^{198}$Hg is qualitatively similar but the mass asymmetry along aEF is significantly reduced, and the energy difference between aEF and sEF (corresponding to $Q_{30}=0$ in Fig.~\ref{fig2}) is small, i.e., the PES is  soft in the octupole direction as one approaches the saddle.
In HFB-D1S calculations of Fig.~\ref{fig1}(d), the energy  balance between these two configurations is so fragile that a local transition from aEF to sEF, and back, is predicted at  large deformations.
This topography is consistent with a broad mass distribution of fission fragments observed in this nucleus. It is interesting to see that the magic structure of $^{90}$Zr manifests itself in a very low energy of sFu in $^{180}$Hg
($^{90}$Zr+$^{90}$Zr) in both models. The asymmetric fusion valleys aFu become more favored in $^{198}$Hg.

\begin{figure}[htb]
  \includegraphics[width=\columnwidth]{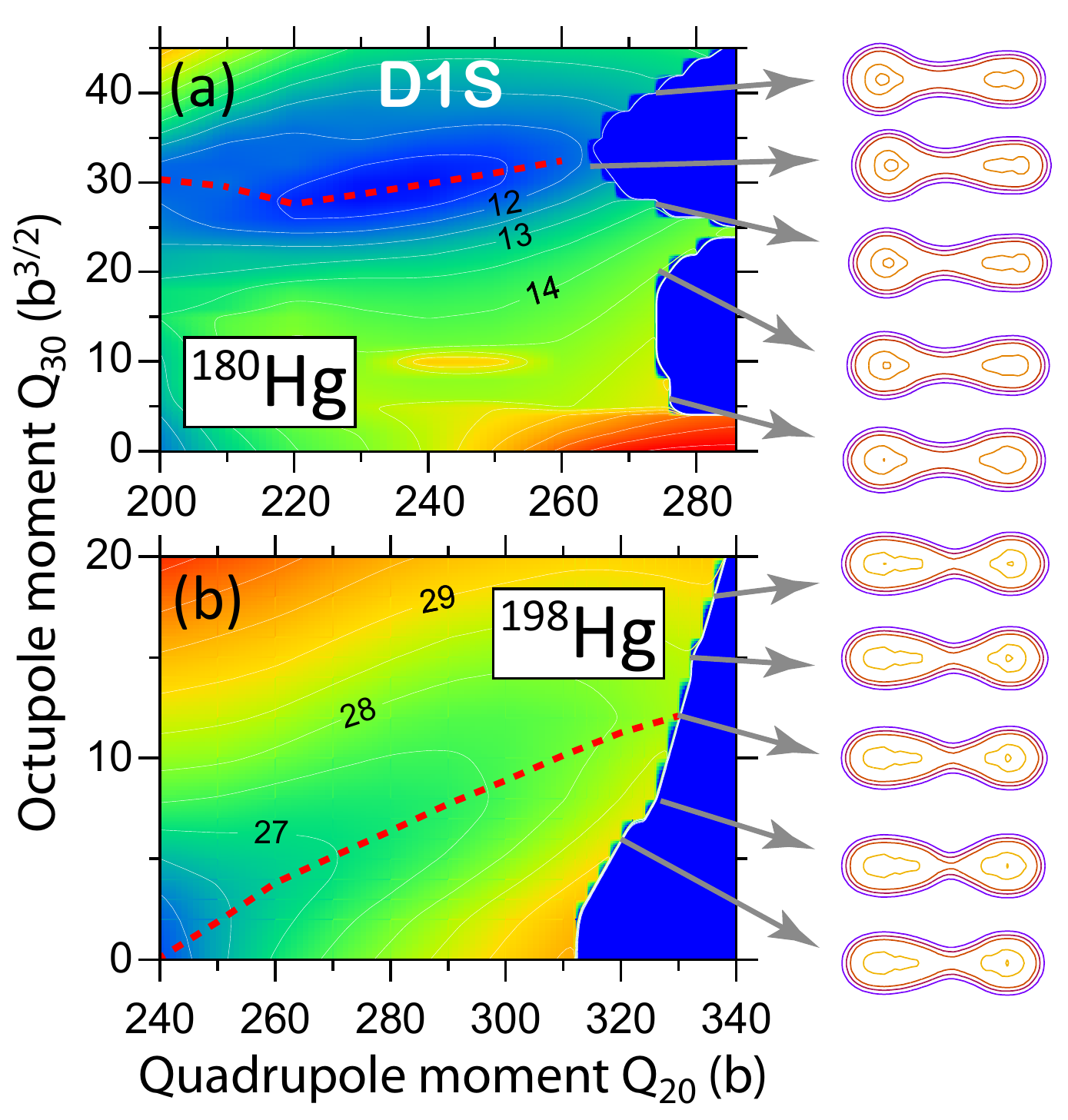}
 \caption[T]{\label{fig3}
  (Color online) PES in HFB-D1S
for $^{180}$Hg (top)  and $^{198}$Hg (bottom)   in the  ($Q_{20}, Q_{30}$) plane in the pre-scission region  of aEF valley. The   symmetric limit corresponds to  $Q_{30}=0$. The aEF valley is marked by a dashed line.  Density profiles for various pre-scission configurations are indicated.
The difference between contour lines is 0.5\,MeV. Note different $Q_{30}$-scales
in   $^{180}$Hg and $^{198}$Hg plots.
 }
\end{figure}
%

%%
%\begin{figure}[htb]
%  \includegraphics[width=\columnwidth]{fig4.pdf}
\begin{figure*}[htb]
  \includegraphics[width=0.8\textwidth]{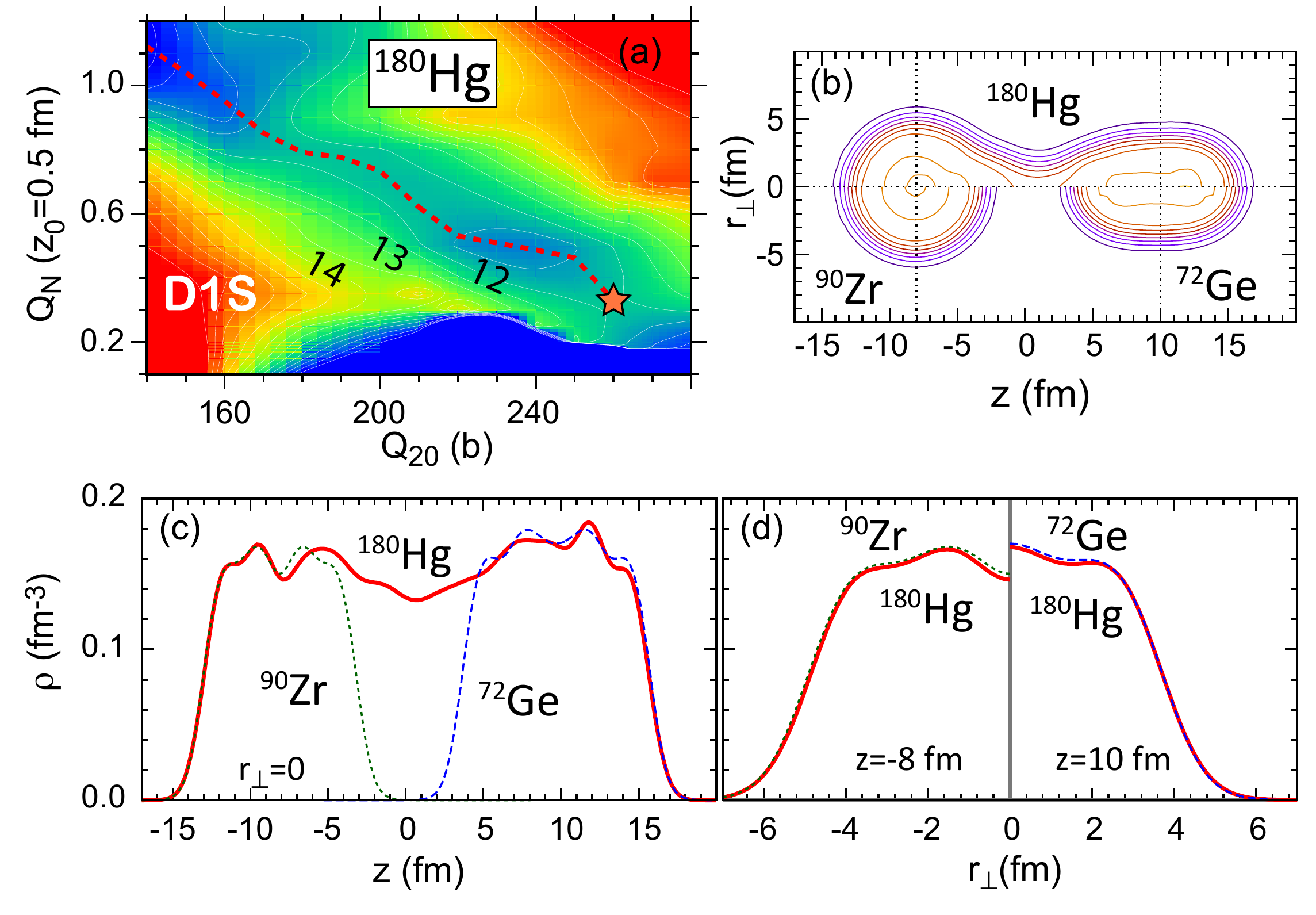}
  \caption[T]{\label{fig4}
  (Color online)  (a) PES of $^{180}$Hg
 in the  ($Q_{20}, Q_N$) plane computed in HFB-D1S in the scission region of  aEF. (b) Density distribution in $^{180}$Hg close to scission at
 $Q_{20}=260$\,b, $Q_N=0.3$, and $Q_{30}=33.8$\,b$^{3/2}$ (marked by a star in panel (a)) compared to density distributions of $^{90}$Zr (in its spherical ground state) and $^{72}$Ge (in the excited deformed configuration with $Q_{20}=8$\,b). The density profiles for $r_\perp=0$  (c), and for $z=-8$\,fm and $z=10$\,fm  (d) along the cuts marked by dotted lines in panel (b).
 }
\end{figure*}
%%
%%%%%%%%%%%%%%%%%%%%%%%%%%%%%%%%%%%%%%%%%%%%%%%%%%%%%%%%
\section{\label{sec:scission}Pre-scission configurations}
%\textit{Pre-scission configurations}---

The properties of fission fragments are governed by the nature of scission configurations at which  a nucleus splits \cite{You11}. 
The scission point is not precisely defined in the models yielding leptodermous densities. We assume that a scission configuration corresponds to a well-defined thin neck, and for greater elongations  the neck decreases  and the binding energy rapidly drops due to the Coulomb repulsion between the fragments. 
For the detailed analysis, the hypersurface of scission
points in the collective space has to be computed \cite{You09}.
To get more insight into the mass distributions of fissioning Hg nuclei, in Fig.~\ref{fig3} we show the topography of PES
for $^{180}$Hg  and $^{198}$Hg  obtained in HFB-D1S for pre-scission configurations around the aEF valley.  It is gratifying to see that the predictions of HFB-D1S and HFB-SkM$^*$ (shown in Fig.~\ref{fig2}) are similar. Namely, in both cases the aEF valley is separated by a ridge from the symmetric $Q_{30}$=0 line (sEF) for $^{180}$Hg, while for  $^{198}$Hg the pre-scission surface is fairly soft in the $Q_{30}$ direction, and the aEF pathway corresponds to much smaller mass asymmetries.

The profiles of the
density distribution for various configurations are also plotted in Fig.~\ref{fig3}.
The density profile corresponding to aEF (the most probable
static scission point) in $^{180}$Hg  can be associated with a $A_H/A_L$=99/81 mass split, which is very consistent with the observed mass asymmetry of $A_H/A_L$=100(1)/80(1) \cite{And10}. The minimum of the neck joining the two pre-fragments
is located  at  0.6\,fm from the center of
mass. Density distributions for higher-lying pre-scission configurations shown in Fig.~\ref{fig3}(a) are all similar, with the heavier fragment being nearly spherical and  the lighter fragment elongated. When moving closer to the $Q_{30}=0$ line, we see that the neck is still pronounced and the fragments are elongated. That is, the symmetric configurations competing with aEF do not correspond to the
two spherical $^{90}$Zr nuclei. A very similar situation is obtained in HFB-SkM$^*$, where the predicted mass split at the static scission point is $A_H/A_L$=101/79, and in the model of Ref.~\cite{ich12}:
$A_H/A_L \approx 103/77$.

The pre-scission shapes  of $^{198}$Hg  are shown in Fig.~\ref{fig3}(b).
The mass split along aEF is  108/90  and the minimum of the neck is located  at about 1.7\,fm from the center of mass.
Interestingly, the heavier fragment is elongated while the lighter fragment is nearly spherical, i.e., this is exactly opposite to  what has been predicted for $^{180}$Hg. Again, our HFB-SkM$^*$ calculations yield a very consistent result,
$A_H/A_L$=110/88, that is close to the mass split $\approx$111/87  of Ref.~\cite{ich12}.

Figure~\ref{fig4}(a) shows  the fission valley aEF for $^{180}$Hg in HFB-D1S
in the  ($Q_{20}, Q_N$) plane as the scission point (small values of $Q_N$)
is gradually approached with increasing $Q_{20}$. For
$Q_{20}<170$\,b, the  configuration sFu  corresponding to two separated $^{90}$Zr fragments lies
above aEF (see also Fig.~\ref{fig1}(b)). For $Q_{20}> 170$\,b
an energy barrier appears  between aEF and sFu. Moreover, as discussed above, a transition from aEF to sFu is going to be strongly hindered by the very different intrinsic structure of these two configurations.

\begin{figure*}[t!]
  \includegraphics[width=0.8\textwidth]{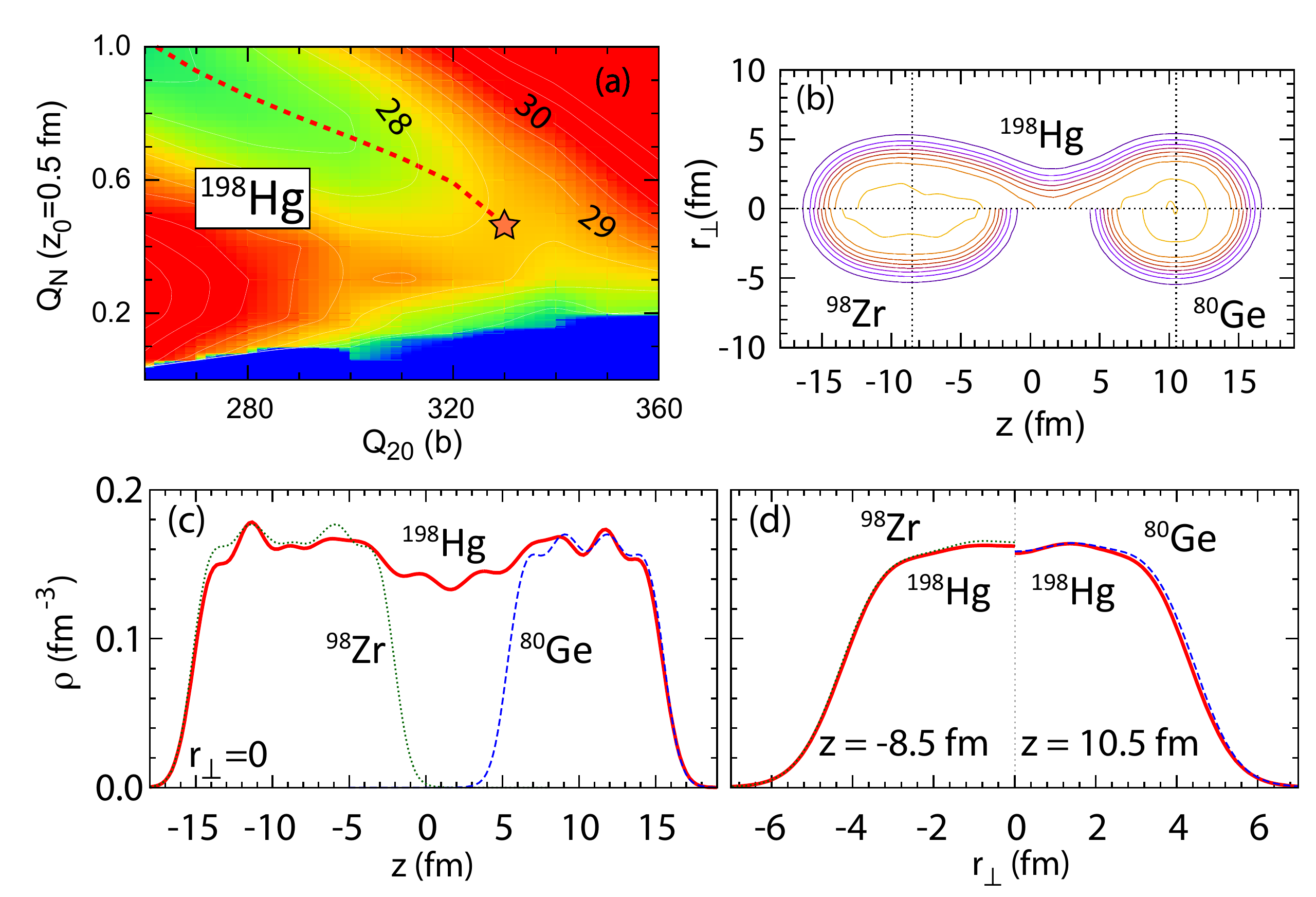}
  \caption[T]{\label{fig5}
  (Color online)  Similar as in Fig.~\ref{fig4} but for $^{198}$Hg.  Density distributions of $^{198}$Hg  in  (b)-(d) correspond to
 $Q_{20}=330$\,b, $Q_N=0.33$, and $Q_{30}=12.1$\,b$^{3/2}$. They are compared to those of  $^{98}$Zr (in the excited deformed configuration with $Q_{20}=12$\,b) and $^{80}$Ge (in its  prolate ground state with $Q_{20}=2.4$\,b). 
 }
\end{figure*}
%%To see what nuclei correspond to the most probable fission split of $^{180}$Hg,
we first considered the nuclei that (i) have the same  $N/Z$ ratio as the parent system; (ii)  have mass numbers  that reproduce the doubled mass of the outer part of the fragment situated outside the vertical line in Fig.~\ref{fig4}(b);  and (iii) have density distributions that match those of the fragments.
The detailed analysis of the shape of $^{180}$Hg in the near-scission configuration of  aEF with  $Q_{20}=260$\,b and $Q_N=0.3$
is shown in  Fig.~\ref{fig4}(b-d). Since
the nuclei $^{90}$Zr and $^{72}$Ge   have the same  $N/Z$ ratio as  $^{180}$Hg
they are obvious candidates.
The larger fragment of  $^{180}$Hg has  a nearly spherical shape. This is consistent with the ground-state of semi-magic $^{90}$Zr. The smaller fragment is strongly elongated. A deformed  configuration of
$^{72}$Ge with  $Q_2=8$\,b, at an excitation energy of 3.4 MeV, fits the bill.
By comparing the calculated density profiles, we see that
the pre-scission configuration
of $^{180}$Hg can indeed be viewed as a molecular system consisting of spherical
$^{90}$Zr and deformed $^{72}$Ge fragments connected by a thin neck with a slightly reduced  density that contains
8 protons and 10 neutrons. At scission
the neck nucleons are shared  between the two fragments. One likely split could be $^{100}$Ru/$^{80}$Kr.

Following similar analysis for $^{198}$Hg presented in Fig.~\ref{fig5}, we conclude that its near-scission aEF configuration can be viewed as a molecular system. It consists of   $^{80}$Ge in its nearly-spherical ground state with $Q_{20}=2.4$\,b,  $^{98}$Zr in a well-deformed  ($Q_{20}=12$\,b) configuration, and a neck
containing  8 protons and 12 neutrons. Therefore, one likely split could be
$^{108}$Ru/$^{90}$Kr.

%%%%%%%%%%%%%%%%%%%%%%%%%%%%%%%%%%%%%%%%%%%%%%%%%%%%%%%%
\section{\label{sec:conclusions}Conclusions}
%
%\textit{Conclusions}---
Our self-consistent calculations based on the nuclear DFT with SkM$^*$ and D1S effective interactions give a very consistent description of the fission process in $^{180}$Hg and $^{198}$Hg. By considering several collective coordinates, we were able to follow  static fission and fusion pathways in the configuration space. We confirm the findings of Ref.~\cite{And10} that the asymmetric fission valley aEF is well separated from  fusion valleys associated with nearly spherical fragments.
We conclude that the mass distribution of fission fragments in both nuclei is governed by shell structure of pre-scission configurations associated with molecular structures. In $^{180}$Hg,  both our models suggest $^{100}$Ru/$^{80}$Kr as the most probable split -- a finding that is very consistent with experiment --  and  both predict symmetric elongated configurations sEF to lie rather high in energy.
The most likely split predicted for  $^{198}$Hg  is  $^{108}$Ru/$^{90}$Kr in HFB-D1S and $^{110}$Ru/$^{88}$Kr in HFB-SkM$^*$.
Both models yield  PES for this nucleus to be fairly soft in the  $Q_{30}$ direction in a pre-scission region, and this is expected to result in an increased yield of nearly-symmetric partitions, and yield a very shallow, or even two-humped, structure seen experimentally \cite{Itk90,Mul98}.

%%%%%%%%%%%%%%%%%%%%%%%%%%%%%%%%%%%%%%%%%%%%%%%%%%%%%%%%
\begin{acknowledgments}
Useful discussions with A.N. Andreyev
are gratefully acknowledged. This work was
supported by
 Grant No. DEC-2011/01/B/ST2/03667 from the National Science Centre (Poland) and
the U.S. Department of Energy under
Contract Nos.\ DE-FC02-09ER41583 (UNEDF SciDAC Collaboration), DE-FG02-96ER40963
 (University of Tennessee),
DE-FG52-09NA29461 (the Stewardship Science Academic Alliances program), and  DE-AC07-05ID14517 (NEUP grant sub award 00091100).
\end{acknowledgments}

%\nocite{*}
%
%\bibliographystyle{apsrev4-1}
%\bibliography{fission}    %Produces the bibliography via BibTeX.

\begin{thebibliography}{34}%
\makeatletter
\providecommand \@ifxundefined [1]{%
 \@ifx{#1\undefined}
}%
\providecommand \@ifnum [1]{%
 \ifnum #1\expandafter \@firstoftwo
 \else \expandafter \@secondoftwo
 \fi
}%
\providecommand \@ifx [1]{%
 \ifx #1\expandafter \@firstoftwo
 \else \expandafter \@secondoftwo
 \fi
}%
\providecommand \natexlab [1]{#1}%
\providecommand \enquote  [1]{``#1''}%
\providecommand \bibnamefont  [1]{#1}%
\providecommand \bibfnamefont [1]{#1}%
\providecommand \citenamefont [1]{#1}%
\providecommand \href@noop [0]{\@secondoftwo}%
\providecommand \href [0]{\begingroup \@sanitize@url \@href}%
\providecommand \@href[1]{\@@startlink{#1}\@@href}%
\providecommand \@@href[1]{\endgroup#1\@@endlink}%
\providecommand \@sanitize@url [0]{\catcode `\\12\catcode `\$12\catcode
  `\&12\catcode `\#12\catcode `\^12\catcode `\_12\catcode `\%12\relax}%
\providecommand \@@startlink[1]{}%
\providecommand \@@endlink[0]{}%
\providecommand \url  [0]{\begingroup\@sanitize@url \@url }%
\providecommand \@url [1]{\endgroup\@href {#1}{\urlprefix }}%
\providecommand \urlprefix  [0]{URL }%
\providecommand \Eprint [0]{\href }%
\providecommand \doibase [0]{http://dx.doi.org/}%
\providecommand \selectlanguage [0]{\@gobble}%
\providecommand \bibinfo  [0]{\@secondoftwo}%
\providecommand \bibfield  [0]{\@secondoftwo}%
\providecommand \translation [1]{[#1]}%
\providecommand \BibitemOpen [0]{}%
\providecommand \bibitemStop [0]{}%
\providecommand \bibitemNoStop [0]{.\EOS\space}%
\providecommand \EOS [0]{\spacefactor3000\relax}%
\providecommand \BibitemShut  [1]{\csname bibitem#1\endcsname}%
\let\auto@bib@innerbib\@empty
%</preamble>
\bibitem [{\citenamefont {Andreyev}\ \emph {et~al.}(2010)\citenamefont
  {Andreyev} \emph {et~al.}}]{And10}%
  \BibitemOpen
  \bibfield  {author} {\bibinfo {author} {\bibfnamefont {A.~N.}\ \bibnamefont
  {Andreyev}} \emph {et~al.},\ }\href {\doibase 10.1103/PhysRevLett.105.252502}
  {\bibfield  {journal} {\bibinfo  {journal} {Phys. Rev. Lett.}\ }\textbf
  {\bibinfo {volume} {105}},\ \bibinfo {pages} {252502} (\bibinfo {year}
  {2010})}\BibitemShut {NoStop}%
\bibitem [{\citenamefont {Krappe}\ and\ \citenamefont
  {Pomorski}(2012)}]{Kra12}%
  \BibitemOpen
  \bibfield  {author} {\bibinfo {author} {\bibfnamefont {H.~J.}\ \bibnamefont
  {Krappe}}\ and\ \bibinfo {author} {\bibfnamefont {K.}~\bibnamefont
  {Pomorski}},\ }\href@noop {} {\emph {\bibinfo {title} {Theory of Nuclear
  Fission: A Textbook}}},\ \bibinfo {series} {Lecture Notes in Physics}, Vol.\
  \bibinfo {volume} {838}\ (\bibinfo  {publisher} {Springer},\ \bibinfo {year}
  {2012})\BibitemShut {NoStop}%
\bibitem [{\citenamefont {Warda}\ \emph {et~al.}(2002)\citenamefont {Warda},
  \citenamefont {Egido}, \citenamefont {Robledo},\ and\ \citenamefont
  {Pomorski}}]{War02}%
  \BibitemOpen
  \bibfield  {author} {\bibinfo {author} {\bibfnamefont {M.}~\bibnamefont
  {Warda}}, \bibinfo {author} {\bibfnamefont {J.~L.}\ \bibnamefont {Egido}},
  \bibinfo {author} {\bibfnamefont {L.~M.}\ \bibnamefont {Robledo}}, \ and\
  \bibinfo {author} {\bibfnamefont {K.}~\bibnamefont {Pomorski}},\ }\href
  {\doibase 10.1103/PhysRevC.66.014310} {\bibfield  {journal} {\bibinfo
  {journal} {Phys. Rev. C}\ }\textbf {\bibinfo {volume} {66}},\ \bibinfo
  {pages} {014310} (\bibinfo {year} {2002})}\BibitemShut {NoStop}%
\bibitem [{\citenamefont {Goutte}\ \emph {et~al.}(2005)\citenamefont {Goutte},
  \citenamefont {Berger}, \citenamefont {Casoli},\ and\ \citenamefont
  {Gogny}}]{Gou05}%
  \BibitemOpen
  \bibfield  {author} {\bibinfo {author} {\bibfnamefont {H.}~\bibnamefont
  {Goutte}}, \bibinfo {author} {\bibfnamefont {J.~F.}\ \bibnamefont {Berger}},
  \bibinfo {author} {\bibfnamefont {P.}~\bibnamefont {Casoli}}, \ and\ \bibinfo
  {author} {\bibfnamefont {D.}~\bibnamefont {Gogny}},\ }\href {\doibase
  10.1103/PhysRevC.71.024316} {\bibfield  {journal} {\bibinfo  {journal} {Phys.
  Rev. C}\ }\textbf {\bibinfo {volume} {71}},\ \bibinfo {pages} {024316}
  (\bibinfo {year} {2005})}\BibitemShut {NoStop}%
\bibitem [{\citenamefont {Staszczak}\ \emph {et~al.}(2009)\citenamefont
  {Staszczak}, \citenamefont {Baran}, \citenamefont {Dobaczewski},\ and\
  \citenamefont {Nazarewicz}}]{Sta09}%
  \BibitemOpen
  \bibfield  {author} {\bibinfo {author} {\bibfnamefont {A.}~\bibnamefont
  {Staszczak}}, \bibinfo {author} {\bibfnamefont {A.}~\bibnamefont {Baran}},
  \bibinfo {author} {\bibfnamefont {J.}~\bibnamefont {Dobaczewski}}, \ and\
  \bibinfo {author} {\bibfnamefont {W.}~\bibnamefont {Nazarewicz}},\ }\href
  {\doibase 10.1103/PhysRevC.80.014309} {\bibfield  {journal} {\bibinfo
  {journal} {Phys. Rev. C}\ }\textbf {\bibinfo {volume} {80}},\ \bibinfo
  {pages} {014309} (\bibinfo {year} {2009})}\BibitemShut {NoStop}%
\bibitem [{\citenamefont {Warda}\ \emph {et~al.}(2005)\citenamefont {Warda},
  \citenamefont {Pomorski}, \citenamefont {Egido},\ and\ \citenamefont
  {Robledo}}]{war05}%
  \BibitemOpen
  \bibfield  {author} {\bibinfo {author} {\bibfnamefont {M.}~\bibnamefont
  {Warda}}, \bibinfo {author} {\bibfnamefont {K.}~\bibnamefont {Pomorski}},
  \bibinfo {author} {\bibfnamefont {J.~L.}\ \bibnamefont {Egido}}, \ and\
  \bibinfo {author} {\bibfnamefont {L.~M.}\ \bibnamefont {Robledo}},\
  }\href@noop {} {\bibfield  {journal} {\bibinfo  {journal} {Int. J. Mod. Phys.
  E}\ }\textbf {\bibinfo {volume} {14}},\ \bibinfo {pages} {403} (\bibinfo
  {year} {2005})}\BibitemShut {NoStop}%
\bibitem [{\citenamefont {Younes}\ and\ \citenamefont {Gogny}(2009)}]{You09}%
  \BibitemOpen
  \bibfield  {author} {\bibinfo {author} {\bibfnamefont {W.}~\bibnamefont
  {Younes}}\ and\ \bibinfo {author} {\bibfnamefont {D.}~\bibnamefont {Gogny}},\
  }\href {\doibase 10.1103/PhysRevC.80.054313} {\bibfield  {journal} {\bibinfo
  {journal} {Phys. Rev. C}\ }\textbf {\bibinfo {volume} {80}},\ \bibinfo
  {pages} {054313} (\bibinfo {year} {2009})}\BibitemShut {NoStop}%
\bibitem [{\citenamefont {Kortelainen}\ \emph {et~al.}(2012)\citenamefont
  {Kortelainen}, \citenamefont {McDonnell}, \citenamefont {Nazarewicz},
  \citenamefont {Reinhard}, \citenamefont {Sarich}, \citenamefont {Schunck},
  \citenamefont {Stoitsov},\ and\ \citenamefont {Wild}}]{Kor12}%
  \BibitemOpen
  \bibfield  {author} {\bibinfo {author} {\bibfnamefont {M.}~\bibnamefont
  {Kortelainen}}, \bibinfo {author} {\bibfnamefont {J.}~\bibnamefont
  {McDonnell}}, \bibinfo {author} {\bibfnamefont {W.}~\bibnamefont
  {Nazarewicz}}, \bibinfo {author} {\bibfnamefont {P.-G.}\ \bibnamefont
  {Reinhard}}, \bibinfo {author} {\bibfnamefont {J.}~\bibnamefont {Sarich}},
  \bibinfo {author} {\bibfnamefont {N.}~\bibnamefont {Schunck}}, \bibinfo
  {author} {\bibfnamefont {M.~V.}\ \bibnamefont {Stoitsov}}, \ and\ \bibinfo
  {author} {\bibfnamefont {S.~M.}\ \bibnamefont {Wild}},\ }\href
  {http://prc.aps.org/abstract/PRC/v85/i2/e024304} {\bibfield  {journal}
  {\bibinfo  {journal} {Phys. Rev. C}\ }\textbf {\bibinfo {volume} {85}},\
  \bibinfo {pages} {024304} (\bibinfo {year} {2012})}\BibitemShut {NoStop}%
\bibitem [{\citenamefont {Schmidt}\ \emph {et~al.}(2000)\citenamefont {Schmidt}
  \emph {et~al.}}]{Schmidt2000}%
  \BibitemOpen
  \bibfield  {author} {\bibinfo {author} {\bibfnamefont {K.-H.}\ \bibnamefont
  {Schmidt}} \emph {et~al.},\ }\href@noop {} {\bibfield  {journal} {\bibinfo
  {journal} {Nucl. Phys. A}\ }\textbf {\bibinfo {volume} {665}},\ \bibinfo
  {pages} {221} (\bibinfo {year} {2000})}\BibitemShut {NoStop}%
\bibitem [{\citenamefont {Myers}\ and\ \citenamefont
  {Swiatecki}(1996)}]{Myers1991}%
  \BibitemOpen
  \bibfield  {author} {\bibinfo {author} {\bibfnamefont {W.~D.}\ \bibnamefont
  {Myers}}\ and\ \bibinfo {author} {\bibfnamefont {W.~J.}\ \bibnamefont
  {Swiatecki}},\ }\href@noop {} {\bibfield  {journal} {\bibinfo  {journal}
  {Nucl. Phys. A}\ }\textbf {\bibinfo {volume} {601}},\ \bibinfo {pages} {141}
  (\bibinfo {year} {1996})}\BibitemShut {NoStop}%
\bibitem [{\citenamefont {Mulgin}\ \emph {et~al.}(1998)\citenamefont {Mulgin},
  \citenamefont {Schmidt}, \citenamefont {Grewe},\ and\ \citenamefont
  {Zhdanov}}]{Mul98}%
  \BibitemOpen
  \bibfield  {author} {\bibinfo {author} {\bibfnamefont {S.~I.}\ \bibnamefont
  {Mulgin}}, \bibinfo {author} {\bibfnamefont {K.-H.}\ \bibnamefont {Schmidt}},
  \bibinfo {author} {\bibfnamefont {A.}~\bibnamefont {Grewe}}, \ and\ \bibinfo
  {author} {\bibfnamefont {S.~V.}\ \bibnamefont {Zhdanov}},\ }\href@noop {}
  {\bibfield  {journal} {\bibinfo  {journal} {Nucl. Phys. A}\ }\textbf
  {\bibinfo {volume} {640}},\ \bibinfo {pages} {375} (\bibinfo {year}
  {1998})}\BibitemShut {NoStop}%
\bibitem [{\citenamefont {Itkis}\ \emph {et~al.}(1990)\citenamefont {Itkis}
  \emph {et~al.}}]{Itk90}%
  \BibitemOpen
  \bibfield  {author} {\bibinfo {author} {\bibfnamefont {M.~G.}\ \bibnamefont
  {Itkis}} \emph {et~al.},\ }\href@noop {} {\bibfield  {journal} {\bibinfo
  {journal} {Sov. J. Nucl. Phys.}\ }\textbf {\bibinfo {volume} {52}},\ \bibinfo
  {pages} {601} (\bibinfo {year} {1990})}\BibitemShut {NoStop}%
\bibitem [{\citenamefont {Itkis}\ \emph {et~al.}(1991)\citenamefont {Itkis}
  \emph {et~al.}}]{Itk91}%
  \BibitemOpen
  \bibfield  {author} {\bibinfo {author} {\bibfnamefont {M.~G.}\ \bibnamefont
  {Itkis}} \emph {et~al.},\ }\href@noop {} {\bibfield  {journal} {\bibinfo
  {journal} {Sov. J. Nucl. Phys.}\ }\textbf {\bibinfo {volume} {53}},\ \bibinfo
  {pages} {757} (\bibinfo {year} {1991})}\BibitemShut {NoStop}%
\bibitem [{\citenamefont {Ichikawa}\ \emph {et~al.}(2012)\citenamefont
  {Ichikawa}, \citenamefont {Iwamoto}, \citenamefont {M\"oller},\ and\
  \citenamefont {Sierk}}]{ich12}%
  \BibitemOpen
  \bibfield  {author} {\bibinfo {author} {\bibfnamefont {T.}~\bibnamefont
  {Ichikawa}}, \bibinfo {author} {\bibfnamefont {A.}~\bibnamefont {Iwamoto}},
  \bibinfo {author} {\bibfnamefont {P.}~\bibnamefont {M\"oller}}, \ and\
  \bibinfo {author} {\bibfnamefont {A.~J.}\ \bibnamefont {Sierk}},\ }\href@noop
  {} {\  (\bibinfo {year} {2012})},\ \Eprint {http://arxiv.org/abs/1203.2011v1
  [nucl-th]} {arXiv:1203.2011v1 [nucl-th]} \BibitemShut {NoStop}%
\bibitem [{\citenamefont {M\"oller}\ \emph {et~al.}(2012)\citenamefont
  {M\"oller}, \citenamefont {Randrup},\ and\ \citenamefont {Sierk}}]{Mol12}%
  \BibitemOpen
  \bibfield  {author} {\bibinfo {author} {\bibfnamefont {P.}~\bibnamefont
  {M\"oller}}, \bibinfo {author} {\bibfnamefont {J.}~\bibnamefont {Randrup}}, \
  and\ \bibinfo {author} {\bibfnamefont {A.~J.}\ \bibnamefont {Sierk}},\
  }\href@noop {} {\bibfield  {journal} {\bibinfo  {journal} {Phys. Rev. C}\
  }\textbf {\bibinfo {volume} {85}},\ \bibinfo {pages} {024306} (\bibinfo
  {year} {2012})}\BibitemShut {NoStop}%
\bibitem [{\citenamefont {Karpov}\ \emph {et~al.}(2008)\citenamefont {Karpov},
  \citenamefont {Keli{\'c}},\ and\ \citenamefont {Schmidt}}]{Kar08}%
  \BibitemOpen
  \bibfield  {author} {\bibinfo {author} {\bibfnamefont {A.~V.}\ \bibnamefont
  {Karpov}}, \bibinfo {author} {\bibfnamefont {A.}~\bibnamefont {Keli{\'c}}}, \
  and\ \bibinfo {author} {\bibfnamefont {K.-H.}\ \bibnamefont {Schmidt}},\
  }\href@noop {} {\bibfield  {journal} {\bibinfo  {journal} {J. Phys. G}\
  }\textbf {\bibinfo {volume} {35}},\ \bibinfo {pages} {035104} (\bibinfo
  {year} {2008})}\BibitemShut {NoStop}%
\bibitem [{\citenamefont {Staszczak}\ \emph {et~al.}(2010)\citenamefont
  {Staszczak}, \citenamefont {Stoitsov}, \citenamefont {Baran},\ and\
  \citenamefont {Nazarewicz}}]{Sta10}%
  \BibitemOpen
  \bibfield  {author} {\bibinfo {author} {\bibfnamefont {A.}~\bibnamefont
  {Staszczak}}, \bibinfo {author} {\bibfnamefont {M.}~\bibnamefont {Stoitsov}},
  \bibinfo {author} {\bibfnamefont {A.}~\bibnamefont {Baran}}, \ and\ \bibinfo
  {author} {\bibfnamefont {W.}~\bibnamefont {Nazarewicz}},\ }\href {\doibase
  10.1140/epja/i2010-11018-9} {\bibfield  {journal} {\bibinfo  {journal} {Eur.
  Phys. J. A}\ }\textbf {\bibinfo {volume} {46}},\ \bibinfo {pages} {85}
  (\bibinfo {year} {2010})}\BibitemShut {NoStop}%
\bibitem [{\citenamefont {Staszczak}\ \emph {et~al.}(2011)\citenamefont
  {Staszczak}, \citenamefont {Baran},\ and\ \citenamefont
  {Nazarewicz}}]{Sta11}%
  \BibitemOpen
  \bibfield  {author} {\bibinfo {author} {\bibfnamefont {A.}~\bibnamefont
  {Staszczak}}, \bibinfo {author} {\bibfnamefont {A.}~\bibnamefont {Baran}}, \
  and\ \bibinfo {author} {\bibfnamefont {W.}~\bibnamefont {Nazarewicz}},\
  }\href {\doibase 10.1142/S0218301311017995} {\bibfield  {journal} {\bibinfo
  {journal} {Int. J. Mod. Phys. E}\ }\textbf {\bibinfo {volume} {20}},\
  \bibinfo {pages} {552} (\bibinfo {year} {2011})}\BibitemShut {NoStop}%
\bibitem [{\citenamefont {Schunck}\ \emph {et~al.}(2012)\citenamefont
  {Schunck}, \citenamefont {Dobaczewski}, \citenamefont {McDonnell},
  \citenamefont {Satu{\l}a}, \citenamefont {Sheikh}, \citenamefont {Staszczak},
  \citenamefont {Stoitsov},\ and\ \citenamefont {Toivanen}}]{Sch11}%
  \BibitemOpen
  \bibfield  {author} {\bibinfo {author} {\bibfnamefont {N.}~\bibnamefont
  {Schunck}}, \bibinfo {author} {\bibfnamefont {J.}~\bibnamefont
  {Dobaczewski}}, \bibinfo {author} {\bibfnamefont {J.}~\bibnamefont
  {McDonnell}}, \bibinfo {author} {\bibfnamefont {W.}~\bibnamefont
  {Satu{\l}a}}, \bibinfo {author} {\bibfnamefont {J.~A.}\ \bibnamefont
  {Sheikh}}, \bibinfo {author} {\bibfnamefont {A.}~\bibnamefont {Staszczak}},
  \bibinfo {author} {\bibfnamefont {M.}~\bibnamefont {Stoitsov}}, \ and\
  \bibinfo {author} {\bibfnamefont {P.}~\bibnamefont {Toivanen}},\ }\href
  {\doibase 10.1016/j.cpc.2011.08.013} {\bibfield  {journal} {\bibinfo
  {journal} {Comput. Phys. Commun.}\ }\textbf {\bibinfo {volume} {183}},\
  \bibinfo {pages} {166} (\bibinfo {year} {2012})}\BibitemShut {NoStop}%
\bibitem [{\citenamefont {Bartel}\ \emph {et~al.}(1982)\citenamefont {Bartel},
  \citenamefont {Quentin}, \citenamefont {Brack}, \citenamefont {Guet},\ and\
  \citenamefont {H{\aa}kansson}}]{Bar82}%
  \BibitemOpen
  \bibfield  {author} {\bibinfo {author} {\bibfnamefont {J.}~\bibnamefont
  {Bartel}}, \bibinfo {author} {\bibfnamefont {P.}~\bibnamefont {Quentin}},
  \bibinfo {author} {\bibfnamefont {M.}~\bibnamefont {Brack}}, \bibinfo
  {author} {\bibfnamefont {C.}~\bibnamefont {Guet}}, \ and\ \bibinfo {author}
  {\bibfnamefont {H.-B.}\ \bibnamefont {H{\aa}kansson}},\ }\href {\doibase
  10.1016/0375-9474(82)90403-1} {\bibfield  {journal} {\bibinfo  {journal}
  {Nucl. Phys. A}\ }\textbf {\bibinfo {volume} {386}},\ \bibinfo {pages} {79}
  (\bibinfo {year} {1982})}\BibitemShut {NoStop}%
\bibitem [{\citenamefont {Dobaczewski}\ \emph {et~al.}(2002)\citenamefont
  {Dobaczewski}, \citenamefont {Nazarewicz},\ and\ \citenamefont
  {Stoitsov}}]{Dob02}%
  \BibitemOpen
  \bibfield  {author} {\bibinfo {author} {\bibfnamefont {J.}~\bibnamefont
  {Dobaczewski}}, \bibinfo {author} {\bibfnamefont {W.}~\bibnamefont
  {Nazarewicz}}, \ and\ \bibinfo {author} {\bibfnamefont {M.~V.}\ \bibnamefont
  {Stoitsov}},\ }\href@noop {} {\bibfield  {journal} {\bibinfo  {journal} {Eur.
  Phys. J. A}\ }\textbf {\bibinfo {volume} {15}},\ \bibinfo {pages} {21}
  (\bibinfo {year} {2002})}\BibitemShut {NoStop}%
\bibitem [{\citenamefont {Dobaczewski}\ \emph {et~al.}(1984)\citenamefont
  {Dobaczewski}, \citenamefont {Flocard},\ and\ \citenamefont
  {Treiner}}]{Dob84}%
  \BibitemOpen
  \bibfield  {author} {\bibinfo {author} {\bibfnamefont {J.}~\bibnamefont
  {Dobaczewski}}, \bibinfo {author} {\bibfnamefont {H.}~\bibnamefont
  {Flocard}}, \ and\ \bibinfo {author} {\bibfnamefont {J.}~\bibnamefont
  {Treiner}},\ }\href@noop {} {\bibfield  {journal} {\bibinfo  {journal} {Nucl.
  Phys. A}\ }\textbf {\bibinfo {volume} {422}} (\bibinfo {year}
  {1984})}\BibitemShut {NoStop}%
\bibitem [{\citenamefont {Warda}\ and\ \citenamefont {Robledo}(2011)}]{War11}%
  \BibitemOpen
  \bibfield  {author} {\bibinfo {author} {\bibfnamefont {M.}~\bibnamefont
  {Warda}}\ and\ \bibinfo {author} {\bibfnamefont {L.~M.}\ \bibnamefont
  {Robledo}},\ }\href@noop {} {\bibfield  {journal} {\bibinfo  {journal} {Phys.
  Rev. C}\ }\textbf {\bibinfo {volume} {84}},\ \bibinfo {pages} {044608}
  (\bibinfo {year} {2011})}\BibitemShut {NoStop}%
\bibitem [{L. M. Robledo, HFBaxial code()}]{axialcode}%
  \BibitemOpen
  L. M. Robledo, HFBaxial code,\ \href@noop {} {} (\bibinfo {year}
  {2002})\BibitemShut {NoStop}%
\bibitem [{\citenamefont {Egido}\ \emph {et~al.}(1997)\citenamefont {Egido},
  \citenamefont {Robledo},\ and\ \citenamefont {Chasman}}]{Egi97}%
  \BibitemOpen
  \bibfield  {author} {\bibinfo {author} {\bibfnamefont {J.~L.}\ \bibnamefont
  {Egido}}, \bibinfo {author} {\bibfnamefont {L.~M.}\ \bibnamefont {Robledo}},
  \ and\ \bibinfo {author} {\bibfnamefont {R.~R.}\ \bibnamefont {Chasman}},\
  }\href@noop {} {\bibfield  {journal} {\bibinfo  {journal} {Phys. Lett. B}\
  }\textbf {\bibinfo {volume} {393}},\ \bibinfo {pages} {13} (\bibinfo {year}
  {1997})}\BibitemShut {NoStop}%
\bibitem [{\citenamefont {Robledo}\ and\ \citenamefont
  {Bertsch}(2011)}]{rob11}%
  \BibitemOpen
  \bibfield  {author} {\bibinfo {author} {\bibfnamefont {L.~M.}\ \bibnamefont
  {Robledo}}\ and\ \bibinfo {author} {\bibfnamefont {G.~F.}\ \bibnamefont
  {Bertsch}},\ }\href {\doibase 10.1103/PhysRevC.84.014312} {\bibfield
  {journal} {\bibinfo  {journal} {Phys. Rev. C}\ }\textbf {\bibinfo {volume}
  {84}},\ \bibinfo {pages} {014312} (\bibinfo {year} {2011})}\BibitemShut
  {NoStop}%
\bibitem [{\citenamefont {Berger}\ \emph {et~al.}(1984)\citenamefont {Berger},
  \citenamefont {Girod},\ and\ \citenamefont {Gogny}}]{ber84}%
  \BibitemOpen
  \bibfield  {author} {\bibinfo {author} {\bibfnamefont {J.~F.}\ \bibnamefont
  {Berger}}, \bibinfo {author} {\bibfnamefont {M.}~\bibnamefont {Girod}}, \
  and\ \bibinfo {author} {\bibfnamefont {D.}~\bibnamefont {Gogny}},\ }\href
  {\doibase 10.1016/0375-9474(84)90240-9} {\bibfield  {journal} {\bibinfo
  {journal} {Nucl. Phys. A}\ }\textbf {\bibinfo {volume} {428}},\ \bibinfo
  {pages} {23} (\bibinfo {year} {1984})}\BibitemShut {NoStop}%
\bibitem [{\citenamefont {M\"oller}\ \emph {et~al.}(2001)\citenamefont
  {M\"oller}, \citenamefont {Madland}, \citenamefont {Sierk},\ and\
  \citenamefont {Iwamoto}}]{Mol01}%
  \BibitemOpen
  \bibfield  {author} {\bibinfo {author} {\bibfnamefont {P.}~\bibnamefont
  {M\"oller}}, \bibinfo {author} {\bibfnamefont {D.~G.}\ \bibnamefont
  {Madland}}, \bibinfo {author} {\bibfnamefont {A.~J.}\ \bibnamefont {Sierk}},
  \ and\ \bibinfo {author} {\bibfnamefont {A.}~\bibnamefont {Iwamoto}},\ }\href
  {\doibase 10.1038/35057204} {\bibfield  {journal} {\bibinfo  {journal}
  {Nature}\ }\textbf {\bibinfo {volume} {409}},\ \bibinfo {pages} {785}
  (\bibinfo {year} {2001})}\BibitemShut {NoStop}%
\bibitem [{\citenamefont {Dubray}\ and\ \citenamefont {Regnier}(2011)}]{Dub11}%
  \BibitemOpen
  \bibfield  {author} {\bibinfo {author} {\bibfnamefont {N.}~\bibnamefont
  {Dubray}}\ and\ \bibinfo {author} {\bibfnamefont {D.}~\bibnamefont
  {Regnier}},\ }  \href
  {\doibase 10.1016/j.cpc.2012.05.001} {\bibfield  {journal} {\bibinfo  {journal}
  {Comput. Phys. Commun.}\ }\textbf {\bibinfo {volume} {183}},\ \bibinfo {pages} {2035}
  (\bibinfo {year} {2012})}\BibitemShut {NoStop}%
\bibitem [{\citenamefont {Younes}\ and\ \citenamefont {Gogny}(2011)}]{You11}%
  \BibitemOpen
  \bibfield  {author} {\bibinfo {author} {\bibfnamefont {W.}~\bibnamefont
  {Younes}}\ and\ \bibinfo {author} {\bibfnamefont {D.}~\bibnamefont {Gogny}},\
  }\href {\doibase 10.1103/PhysRevLett.107.132501} {\bibfield  {journal}
  {\bibinfo  {journal} {Phys. Rev. Lett.}\ }\textbf {\bibinfo {volume} {107}},\
  \bibinfo {pages} {132501} (\bibinfo {year} {2011})}\BibitemShut {NoStop}%
\end{thebibliography}

\end{document}